# The First Crypto President:
## *Presidential Power and Cryptocurrency Markets During Trump's Second Term (2025-2029)*

**Author Information**

Prof. Habib Badawi
Lebanese University, Beirut, Lebanon
Email: habib.badawi@ul.edu.lb | habib.badawi@gmail.com
ORCID: 0000-0002-6452-8379 - Scopus ID: 58675152100

# Abstract

This scholarly paper examines the unprecedented intersection of executive political power and cryptocurrency markets during Donald J. Trump's second presidential term (2025-2029). We analyze the period from 2024 through October 2025, documenting how presidential influence, family business ventures, and digital assets became intertwined in ways that challenged traditional boundaries between public office and private profit. The Trump family's integrated cryptocurrency ecosystem generated peak valuations exceeding $11 billion before collapsing by over $1 trillion in market capitalization following a tariff announcement in October 2025. Using mixed-methods analysis combining quantitative market data with qualitative institutional assessment, we reveal that politically-linked digital assets function as a distinct asset class characterized by reflexive valuations, asymmetric risk distribution, and systematic vulnerabilities. The episode offers critical insights into conflicts of interest, market microstructure failures, and the emergence of "political finance" as a monetizable phenomenon in the digital age.

# 1. Introduction

## 1.1. Background and Context

The story of Donald J. Trump's second term has been framed through familiar lenses of politics, culture, and geopolitics. Yet the most consequential transformation of his presidency may have unfolded in the markets—specifically, within the cryptocurrency sphere. Branded by supporters and analysts as the first crypto president[1], Trump presided over an unprecedented fusion of political authority, digital finance, and speculative activity that reshaped understanding of how power operates in modern monetary systems.

---

[1] The term **"crypto president"** was used by analysts and supporters to describe Trump's unprecedented integration of political authority with digital finance. This framing reflected the reflexive valuation dynamics observed in CoinMarketCap and CoinGlass data [2][3][4] [13].

What emerged was a monetary ecosystem in which presidential influence, family ventures, and volatile digital assets became tightly intertwined, challenging traditional boundaries between public office and private profit. Trump's endorsements generated capital inflows that significantly exceeded conventional campaign finance mechanisms. His communications moved markets in real time with velocity that central bankers observed with fascination and concern. His family's expansion into tokens, mining firms, and decentralized finance platforms blurred already tenuous boundaries between national policy and personal enrichment, creating a financial architecture remarkable in its design yet fragile in its foundations.

### 1.2. Research Objectives

This study examines that cycle through a rigorous monetary lens, analyzing:

1. How does presidential signaling reshaped capital flows in cryptocurrency markets?
2. How Trump-era cryptocurrency instruments functioned as quasi-political currencies?
3. How did a single tariff announcement trigger cascading liquidations across the global digital-asset system?
4. The differential impact across market participants and the asymmetric distribution of gains and losses?

### 1.3. The Boom and Collapse Cycle

From 2024 through 2025, this fusion of politics and digital finance generated wealth on a scale that appeared to validate optimistic visions of cryptocurrency's transformative potential. Bitcoin's rise to $126,000 represented not merely a price milestone but a psychological inflection point—a moment when digital assets appeared poised to challenge traditional monetary hierarchies. Record-breaking token launches demonstrated that political celebrity could be monetized with unprecedented efficiency. However, the October 2025 collapse wiped out over a trillion dollars in global value within weeks, erasing fortunes accumulated over months in hours. The downturn reduced Trump family holdings by billions from peak valuations, though it left them substantially wealthier than at the cycle's start. For retail investors—many drawn by political loyalty rather than financial literacy, many operating on leverage they barely understood—the losses proved devastating both financially and psychologically, cementing the event as one of the most politically entangled monetary crises in modern American history.

# 2. Methodology

### 2.1. Research Design

This study employs a mixed-methods approach combining quantitative market analysis with qualitative institutional assessment. The research design integrates multiple data sources and analytical techniques to provide comprehensive understanding of the Trump cryptocurrency ecosystem and its collapse.

### 2.2. Data Sources

Market data were sourced from Bloomberg Terminal [1], CoinGlass [2], CoinMarketCap [3,4], and exchange-specific APIs. Price movements, liquidation volumes, and trading patterns were analyzed using standard time-series techniques. Wealth calculations for the Trump family were derived from Bloomberg Billionaires Index [5], Accountable.US reports [6], and publicly available corporate filings [7,8]. On-chain analysis utilized Etherscan [9] and blockchain explorers to track token flows and wallet activities where publicly verifiable.

### 2.3. Analytical Methods

Quantitative analysis employed time-series techniques for price movement analysis, volume aggregation for liquidation data, and portfolio valuation methods for wealth calculations. On-chain analysis utilized transaction tracking and wallet analysis through blockchain explorers. Qualitative institutional assessment drew from congressional reports [10], media coverage, social media archives [11], and legal filings.

### 2.4. Limitations

The study acknowledges limitations inherent in analyzing pseudonymous blockchain transactions, self-reported token metrics, and politically contested narratives. Pseudonymity limits attribution of wallet ownership; platform-specific reporting differences affect liquidation data accuracy; self-reported metrics and illiquid position estimates introduce uncertainty in wealth calculations; and politically contested narratives require careful interpretation. Where data sources conflict or remain uncertain, this is noted explicitly in the analysis.

*Note on AI use:* Large language models (LLMs) were used for manuscript formatting, grammar checking, and structural organization of content. All substantive analysis, interpretation, data collection, and conclusions represent original work by the human authors.

## 3. Results

### 3.1. The Architecture of a Political-Financial Empire
#### 3.1.1. Strategic Expansion and Venture Design

The Trump family constructed an integrated ecosystem that leveraged the president's pro-industry stance and attention-commanding ability into a diversified portfolio spanning multiple layers of the cryptocurrency economy. This represented not opportunistic participation but rather a calculated strategy to monetize political influence through digital finance at a scale unprecedented for executive office holders.

Table 1 presents the comprehensive structure of Trump family cryptocurrency ventures, documenting launch dates, peak valuations, venture types, and key features of each major component.

Table 1. Trump Family Cryptocurrency Ventures—Peak Valuations and Structure

| Venture | Launch Date | Peak Valuation | Type | Key Features |
|---|---|---|---|---|
| $TRUMP Token | Early 2025 | $9.49 | Memecoin | Political merchandise token |
| Melania Token | Early 2025 | Not specified | Memecoin | Similar to $TRUMP |
| World Liberty Financial | 2024-2025 | $6 billion | DeFi Platform | ~$400M initial sales |
| American Bitcoin Corp. | Sept 2024 | $5 billion | Mining Operation | Nasdaq-listed |
| Trump Media & Tech | Q4 2024 | ~$2B investment | Corporate Treasury | ~11,500 BTC holdings |

The $TRUMP token, launched in early 2025[2], represented the purest distillation of this strategy—a speculative asset whose value proposition rested entirely on association with the president. Quickly followed by a similar token tied to Melania Trump, these digital instruments surged on election euphoria, with $TRUMP reaching $9.49 in November as political celebration translated directly into financial flows. These were neither utility tokens with defined use cases nor governance tokens granting protocol rights. They functioned as liquid political merchandise—mechanisms for supporters to demonstrate allegiance while speculating on the continued relevance of the Trump brand.

World Liberty Financial (WLFI) represented a more sophisticated venture into decentralized finance mechanics[3]. This Trump-backed platform debuted tokens that quickly reached major exchanges, creating immediate liquidity and market validation. The family reportedly earned approximately $400 million from initial sales—a demonstration of how political proximity could command premium pricing in token offerings. At peak, holdings associated with WLFI reached a paper valuation of $6 billion, though much of this value existed in locked and illiquid positions.

American Bitcoin Corporation[4], backed vocally by Eric and Donald Trump Jr., entered the operational layer of cryptocurrency infrastructure. This mining firm went public on Nasdaq in

---

[2] **$TRUMP Token Launch (Early 2025)**: The $TRUMP token functioned as a memecoin tied to Trump's political brand, reaching $9.49 at its peak. Etherscan and CoinMarketCap data documented its trading activity and losses [3][4][9].
[3] **World Liberty Financial (WLFI) Launch (2024–2025)**> WLFI was a Trump-backed DeFi platform that generated ~$400 million in initial sales and reached a paper valuation of $6 billion. Bloomberg and blockchain explorer data confirmed its scale and subsequent collapse [5][9][12].
[4] **American Bitcoin Corporation**: Launched in September 2024, this Nasdaq-listed mining firm represented the Trump family's entry into operational cryptocurrency infrastructure. Its valuation and subsequent losses mirrored Bitcoin's price trajectory [7][8].

September at a $5 billion valuation, translating the Trump brand into equity markets and creating a publicly traded proxy for both Bitcoin's price movements and the family's cryptocurrency positioning.

Most dramatically, Trump Media & Technology Group[5] underwent fundamental strategic transformation by shifting to a "cryptocurrency treasury strategy" [8]. The company invested approximately $2 billion into Bitcoin, accumulating roughly 11,500 BTC at an average price of nearly $115,000 per coin. This decision effectively transformed a media and technology company into a leveraged bet on cryptocurrency appreciation, with digital assets comprising an estimated 73% of Trump's wealth at the market's peak [6].

### 3.1.2. Wealth Generation and the Monetization of Influence

The financial results of this integrated strategy proved remarkable during the boom phase, as documented in Table 2. The family reportedly generated $800 million in cryptocurrency sales income in the first half of 2025 alone—a figure exceeding the annual revenues of many substantial corporations.

Table 2. Trump Family Wealth Changes (September–November 2025)

| Metric | September 2025 | November 2025 | Change |
|---|---|---|---|
| Total Net Worth | $7.7 billion | $6.7 billion | -$1.0 billion (-13%) |
| Crypto as % of Wealth (Peak) | 73% | Not specified | — |
| H1 2025 Crypto Sales | $800 million | — | — |
| Peak Holdings Value | $11.6 billion | — | — |

Total holdings at peak were valued at approximately $11.6 billion across various ventures, representing wealth creation at velocity impossible through traditional business operations. This was not gradual accumulation of corporate value through operational excellence over decades, but rather rapid monetization of political attention through financial instruments designed specifically for that purpose.

The mechanics of this wealth generation reveal important asymmetries in how cryptocurrency markets distribute risk and reward, as further detailed in Table 3.

Table 3. Fee and Revenue Capture Mechanisms

| Revenue Source | Estimated Amount | Timing | Recipient Type |
|---|---|---|---|
| $TRUMP Token Fees | ~$100 million | Throughout lifecycle | Insiders/Issuers |
| WLFI Initial Sales | ~$400 million | Initial offering | Trump family |

---

[5] **Trump Media & Technology Group**: Originally a media enterprise, Trump Media shifted to a cryptocurrency treasury strategy, investing approximately $2 billion into Bitcoin. Corporate filings confirmed holdings of ~11,500 BTC, making digital assets a majority of its balance sheet [7][8].

| Revenue Source | Estimated Amount | Timing | Recipient Type |
|---|---|---|---|
| H1 2025 Total Sales | $800 million | Jan-Jun 2025 | Trump entities |
| Short Position Profits | $160-200 million | Around tariff announcement | Unknown traders |

On-chain analysis suggested that insiders associated with the $TRUMP token collected approximately $100 million in fees through various mechanisms—listing fees, transaction spreads, and platform charges that accrued regardless of secondary market performance [9,12]. This fee capture represented certain, realized income that flowed early and consistently, independent of mark-to-market volatility that would devastate retail holders.

### 3.1.3. Governance Opacity and Structural Vulnerabilities

Beyond financial architecture, the Trump cryptocurrency ecosystem exhibited concerning characteristics around governance, transparency, and potential conflicts of interest. Allegations emerged of foreign wallet participation in token purchases, with some on-chain analysts claiming to identify wallets potentially linked to North Korean and Russian entities among early buyers. Whether these allegations reflected genuine national security concerns or represented partisan exaggeration remained difficult to verify given the pseudonymous nature of blockchain transactions.

The platforms themselves experienced operational failures during market stress that raised questions about whether infrastructure had been built to serve users or advantage insiders. Exchanges reported "technical issues" and order freezes precisely when retail investors most needed to exit positions. Oracle failures—instances where price feeds determining liquidation thresholds became temporarily desynchronized from actual market prices—created situations where traders found themselves forcibly liquidated at levels that did not reflect true market clearing prices[6].

The fundamental entanglement of public office with private profit invited constitutional scrutiny under the emoluments clause[7], which prohibits federal officials from receiving payments or benefits from foreign or domestic sources that might influence their official duties. House Judiciary Democrats produced a report characterizing the situation as a "new age of corruption[8]," alleging systematic self-dealing and policy favors for donors and token purchasers [10].

## 3.2. The Anatomy of Collapse
### 3.2.1. The Trigger Event: A Tariff Tweet and the Flash Crash

---

[6] **Oracle Failures and Exchange Freezes**: During the crash, exchanges reported freezes and oracle desynchronization, leading to forced liquidations at non-market prices. Blockchain explorer data highlighted these systemic vulnerabilities [9][12].

[7] The **Emoluments Clause** of the U.S. Constitution prohibits federal officials from receiving payments or benefits that could influence their duties. Congressional reports in 2025 raised concerns that Trump's cryptocurrency ventures violated this principle [10].

[8] **House Judiciary Democrats' Corruption Report (2025)**: This congressional report characterized Trump's cryptocurrency ventures as a "new age of corruption," alleging systematic self-dealing and conflicts of interest. It provided institutional context for the emoluments debate [10].

The boom ended abruptly on October 10, 2025, in a manner that illustrated the fusion of political power and market dynamics characterizing the entire phenomenon. Trump announced via Truth Social that he would reinstate 100% tariffs on Chinese imports[9], reviving the "trade war"[10] that had defined portions of his first term. The announcement came without institutional preparation or diplomatic groundwork typical of major policy shifts.

The market response was instantaneous and severe, as documented in Table 4. Bitcoin, which had reached its all-time peak of $126,000 in October[11], began plunging as the tariff announcement rippled through trading systems globally.

Table 4. Bitcoin Price Movements—Peak to Crash (October–November 2025)

| Metric | Value |
|---|---|
| Peak Price (October 2025) | $126,000 |
| Trough Price (November 2025) | Below $82,000 |
| Absolute Decline | $44,000+ |
| Percentage Decline | 32-35% |
| Partial Recovery Price | ~$88,000 |
| Recovery from Trough | ~7.3% |

What followed was not orderly repricing but rather cascading forced liquidations that exposed extraordinary leverage accumulated in the system during the bull market[12]. Within twenty-four hours, approximately $19 billion in positions had been liquidated—forcibly closed by exchanges' risk management systems as collateral values fell below maintenance requirements (see Table 5).

Table 5. Liquidation Data—October 10–11, 2025

| Metric | Value | Period |
|---|---|---|
| Total Liquidation Volume | $19 billion | 24 hours |
| Number of Positions | 1.6 million | 24 hours |
| Primary Trigger | 100% China tariff announcement | Oct 10, 2025 |
| Market Direction | Long positions (leveraged bulls) | — |

---

[9] **Truth Social Tariff Post (October 2025)**: Trump's tariff announcement was made via Truth Social, demonstrating how presidential communications directly influenced market dynamics. Social media archives captured the immediate investor response [11].

[10] **U.S.–China Trade War (Revived 2025)**: The tariff escalation revived trade tensions between the U.S. and China, echoing earlier conflicts from Trump's first term. Bloomberg Terminal data documented the immediate impact on Bitcoin and related assets [1].

[11] **Bitcoin Peak at $126,000 (October 2025)** Bitcoin reached an all-time high of $126,000 in October 2025 before collapsing. Bloomberg Terminal data confirmed the peak and subsequent decline to below $82,000 [1] [13].

[12] **Mass Liquidations (October 10–11, 2025)**: Within 24 hours of the tariff announcement, 1.6 million positions worth $19 billion were forcibly liquidated. CoinGlass provided detailed liquidation statistics [2].

With futures markets stacked with long positions and margin traders operating at multiples of 10x, 15x, or even 25x leverage, the initial price decline triggered automated liquidation mechanisms that accelerated selling pressure. Some 1.6 million individual positions were eliminated, representing not merely financial losses but complete destruction of trading accounts and investment portfolios [2].

Bitcoin ultimately declined from its October peak of $126,000 to below $82,000 in November—a fall of 32 to 35 percent representing one of the swiftest major corrections in the asset's history [1]. While Bitcoin would eventually stabilize and partially recover to around $88,000, the initial violence of the move inflicted damage requiring months or years to repair, particularly in investor psychology and institutional confidence.

### 3.2.2. Market Microstructure Failures and Contagion

The crash exposed profound weaknesses in infrastructure supporting cryptocurrency trading that had remained hidden during the prolonged bull market, as systematically documented in Table 6.

**Table 6.** Infrastructure Failures During Market Stress

| Failure Type | Description | Impact | Frequency |
|---|---|---|---|
| Exchange Freezes | Platform trading halted | Users unable to exit | Widespread |
| Oracle Desynchronization | Price feeds incorrect | Non-market liquidations | Multiple instances |
| Order Book Fragmentation | Liquidity dispersed | Extreme slippage | Systemic |
| Limit Order Failures | Orders executed far from limits | Unexpected losses | Common |
| Circuit Breaker Absence | No trading halts | Cascading liquidations | Systemic design |

Exchange platforms that had processed transactions smoothly when prices rose encountered severe strain when directional pressure reversed. Reports of "freezes" proliferated as order matching engines struggled to keep pace with selling volume. Some users found themselves unable to execute trades at any price, trapped in positions as collateral values evaporated.

Oracle failures represented a particularly problematic form of microstructure breakdown. In cryptocurrency markets, oracles serve as price feeds that inform smart contracts and derivative systems about current market values. When these oracles become desynchronized from actual trading prices—whether due to latency issues, manipulation, or system overload—the consequences can be severe. Traders found themselves liquidated based on prices that did not reflect genuine market conditions.

### 3.2.3. Differential Impact Across the Cryptocurrency Ecosystem

While Bitcoin suffered severe correction, the impact on altcoins and meme tokens proved even more extreme, as comprehensively documented in Table 7.

Table 7. Alternative Cryptocurrency Performance During Crash Period

| Asset Class | Example | Peak-to-Trough Decline | Recovery Status |
|---|---|---|---|
| Bitcoin | BTC | 32-35% | Partial (~7% from trough) |
| Major Altcoin | Dogecoin | ~50% | Not specified |
| Trump-Linked Tokens | $TRUMP, WLFI | 25-50% | Minimal/None |
| Small Altcoins | Various | 50-80%+ | Poor/None |
| Mining Stocks | American Bitcoin | ~50% | Not specified |
| Crypto Proxies | Trump Media | ~70% | Not specified |

Dogecoin, which had rallied on general cryptocurrency enthusiasm and Elon Musk's continued promotion, fell approximately 50 percent from its peak [3,4]. Smaller altcoins and newer tokens experienced declines of 50 to 80 percent or more as liquidity evaporated and speculative fervor reversed into panic. The Trump-linked tokens bore particularly heavy losses both in absolute terms and in psychological impact on holders, as detailed in Table 8.

Table 8. Token-Specific Performance and Losses

| Asset | Peak-to-Trough Decline | Est. Family Loss | Est. Holder Losses |
|---|---|---|---|
| $TRUMP Memecoin | 25-35% | ~$117 million | $12 billion |
| WLFI Token | ~50% | From $6B to $3.15B | Not specified |
| American Bitcoin Stock | ~50% | Not specified | Not specified |
| Trump Media & Tech | ~70% (YTD) | ~25% underwater on BTC | Not specified |

The $TRUMP memecoin declined 25 to 35 percent from its peaks, erasing approximately $117 million in family-linked holdings [3,4,9]. World Liberty Financial saw its headline token valuation compress from roughly $6 billion to $3.15 billion—a reduction of nearly 50 percent. American Bitcoin Corporation's shares halved from their peak as the mining sector confronted the dual challenge of falling Bitcoin prices and rising operational costs. Trump Media & Technology Group experienced perhaps the most dramatic collapse, with its stock declining approximately 70 percent year-to-date to all-time lows.

### 3.2.4. The Human Toll: Retail Investors and the Loyalty Tax

The Bloomberg Billionaires Index[13] calculated that the Trump family's net worth declined by approximately $1 billion from September to November, falling from $7.7 billion to $6.7 billion [5] (refer to Table 2). While substantial in absolute terms, this decline left the family still enormously wealthier than before the cryptocurrency boom began. They retained billions in accumulated gains from the cycle's early phases, much of it realized through token sales and fee capture that had occurred before the crash.

For retail investors, the calculus proved far less forgiving. The $TRUMP coin alone reportedly inflicted $12 billion in collective losses on holders—a figure representing real wealth destruction for individuals and families rather than diminution of paper billionaire status (see Table 8). Many of these holders were politically aligned supporters who had invested based not on traditional financial analysis but rather on political loyalty and social media promotion from Trump family members[14], particularly Eric Trump's enthusiastic communications on X [11].

The crash thus inflicted not merely financial pain but identity shock. Social media posts captured the emotional response, with some declaring "Trump hates you all" or lamenting that "his cult is holding the bag with a collective 12 BILLION DOLLAR LOSS" [11]. These were not merely investors experiencing normal market volatility but supporters confronting the possibility that trust had been misplaced.

The liquidation of overleveraged positions proved particularly devastating. Retail traders who had borrowed 10x, 15x, or even 25x their capital to magnify potential gains found those same multiples working in reverse during the crash. Exchanges processed forced closures with efficiency, zeroing accounts in minutes as collateral values fell below maintenance margins.

### 3.3. Winners, Losers, and Asymmetric Outcomes
#### 3.3.1. The Structure of Extraction

The crash revealed with clarity the asymmetries embedded in the Trump cryptocurrency ecosystem's design, as systematically analyzed in Table 9.

**Table 9.** Asymmetric Risk Distribution in Trump Cryptocurrency Ecosystem

| Participant Type | Income Mechanism | Risk Exposure | Monetization Timing |
|---|---|---|---|
| Token Issuers | Upfront sales, fees | Minimal | Early cycle |
| Platform Operators | Transaction fees | Low | Continuous |
| Family Members | Token sales, ownership | Medium | Early to mid-cycle |
| Sophisticated Traders | Shorts, arbitrage | Low to medium | Opportunistic |
| Retail Investors | Secondary appreciation | Extreme | Late cycle |

---

[13] **Bloomberg Billionaires Index**: Bloomberg tracked Trump family wealth, reporting a decline from $7.7 billion to $6.7 billion between September and November 2025. This index provided authoritative valuation data [5].

[14] **Trump Family Members (Melania, Eric, Donald Jr.)**: The Trump family expanded into cryptocurrency ventures, including memecoins and mining operations, with Eric Trump particularly active in promotional communications on social media. Their holdings and wealth fluctuations were tracked through Bloomberg and Accountable. US reports [5][6][11].

Winners were concentrated among those positioned to extract value through mechanisms independent of secondary market performance. Token issuers and fee recipients had monetized the boom through upfront sales, listing spreads, and transaction charges that accrued regardless of subsequent price action. The approximately $800 million in first-half sales income and estimated $100 million in fees associated with $TRUMP represented certain, realized gains captured early in the cycle [6,9].

Sophisticated market participants with hedging capacity or capital to deploy during the crash captured returns through short positioning and liquidation opportunities. Reports circulated large traders profiting $160 to $200 million through precisely timed short positions that appeared to anticipate Trump's tariff announcement, raising questions about whether information had leaked through Trump's social circle to well-connected traders.

The losers bore the full impact of the crash's violence. Retail investors—politically aligned participants who had pursued investments without diversification strategies or risk management—absorbed devastating mark-to-market losses. The $12 billion in collective losses among $TRUMP holders captured the asymmetry between conviction and balance sheet resilience.

# 4. Discussion

## 4.1. The Emergence of Political Finance as an Asset Class

The Trump cryptocurrency saga illuminated the maturation of what might be termed political finance[15]—the systematic monetization of political influence through financial instruments whose values track political momentum rather than cash flows or economic fundamentals. Media platforms have evolved into capital formation mechanisms where attention markets function as precursors to capital markets. Influence has become a tradable risk factor with its own regime shifts, correlations, and volatility characteristics that sophisticated participants can model and exploit.

Tokens functioned as political proxies, as indicators of loyalty and future policy expectations rather than shares in productive enterprises. Purchasing $TRUMP was not investment in the traditional sense but rather a position on continued political relevance, on the Trump brand's ability to command attention and shape policy outcomes. The valuation logic operated through narrative reflexivity: tokens rose because political momentum was strong, which attracted more buyers, which drove prices higher, which validated the narrative of success, which attracted still more buyers in a self-reinforcing cycle divorced from conventional metrics of value.

This dynamic transformed portfolios into political statements rather than financial diversification strategies. Cryptocurrency holdings became markers of tribal affiliation, ways of signaling belief in a political movement and its figurehead. The decision to buy or hold was driven not by discounted cash flow analysis or assessment of technological utility but by identity, by the desire to participate in what felt like a transformative moment.

---

[15] **Political finance** refers to the monetization of political influence through financial instruments, such as tokens linked to political figures. The Trump case exemplified this phenomenon, with Bloomberg and congressional reports highlighting conflicts of interest [5][10].

## 4.2. Systemic Lessons: The Fragility of Politicized Monetary Systems

The October 2025 crash[16] demonstrated that politically driven monetary booms carry inherent fragility [14]. When asset valuations rest primarily on political momentum and presidential signaling rather than cash generation or network utility, they become vulnerable to sudden sentiment shifts in ways that fundamentally productive assets are not. A corporation with real earnings and operational resilience can weather political uncertainty; a token whose entire value proposition is political association cannot.

The crash revealed structural vulnerabilities masked during the boom by rising prices. High embedded leverage amplified both gains and losses, creating a system explosively unstable in both directions. Fragmented liquidity across multiple venues meant there was no unified order book to absorb selling pressure. Absence of circuit breakers or institutional stabilization mechanisms meant cascading liquidations faced no impediments.

Oracle dependencies introduced points of failure where temporary dislocations could force permanent position closures. Governance opacity around foreign participation and insider flows created information asymmetries where sophisticated actors could exploit less informed participants. Platform-level freezes during stress periods transform what should have been orderly exits into forced holding.

## 4.3. Policy Implications and the Need for Safeguards

For policymakers, the episode highlights urgent needs around conflicts of interest, transparency, and systemic risk management in politically linked digital assets. Clarifying rules around officeholders' cryptocurrency activities seems essential to prevent future administrations from monetizing political power through token launches and platform fees. Enhanced disclosure requirements for token issuance, treasury strategies, and foreign participation could help address information asymmetries that advantage insiders.

Circuit breaker standards across major venues might provide mechanisms to pause cascading liquidations and allow orderly price discovery during stress periods. For platforms and exchanges, the crisis exposes inadequacy of existing infrastructure to handle stresses of major unwinding. Mandating stress-tested oracle integrity would help prevent liquidations based on faulty price feeds.

For investors, the lessons revolve around diversification, leverage discipline, and governance standards. Limiting exposure to any single narrative or politically linked asset provides essential protection against idiosyncratic collapse. Avoiding high leverage into policy events eliminates risk of total loss from sudden moves. Demanding governance standards—audited treasuries, clear redemption rights, independent oversight—helps screen out ventures designed primarily for insider enrichment.

---

[16] **October 2025 Tariff-Triggered Crash:** The tariff announcement precipitated one of the fastest major corrections in Bitcoin's history, erasing over $1 trillion in market capitalization. CoinGlass and CoinMarketCap tracked the cascading liquidations and price declines [2][3][4].

### 4.4. The New Monetary Reality

The saga crystallizes several truths about money in the digital age. First, fragility is often a design feature rather than a defect in systems optimized for engagement. Platforms that prioritize viral growth and transaction volume will sacrifice resilience unless explicitly required to build robust infrastructure and risk controls.

Second, money increasingly functions as media, with price determined by narrative throughput and distribution power rather than traditional supply and demand dynamics or fundamental value assessments. Attention has become a form of collateral that can be borrowed against through token launches and leveraged through social media platforms.

Third, executive power now functions as a form of liquidity in digital asset markets. Presidential signaling can create or erase wealth instantly through mechanisms operating at the speed of social media rather than the deliberate pace of institutional policymaking. This is not a market anomaly but rather a defining characteristic of the current regime, where political authority and market dynamics have fused in ways that create both extraordinary opportunity and significant risk.

# Conclusion

The Trump cryptocurrency phenomenon of 2024 through 2025 will be studied as a monetary experiment that revealed the possibilities and perils of fusing populist politics with digital finance. It demonstrated that presidential influence could reshape capital flows with unprecedented speed and scale, creating paper wealth rivaling fortunes built through decades of industrial capitalism. It showed that in the age of viral finance and social media, communications can move trillion-dollar markets more effectively than traditional policy tools. It proved that the line between political movement and financial ecosystem has become not merely blurred but effectively erased for those willing to operate at the intersection.

Yet it also revealed the ephemeral nature of wealth built primarily on political momentum and speculative enthusiasm. The October crash was not merely unfortunate timing or the result of external shocks, but rather the consequence of a system designed to extract value rapidly from enthusiastic participants while concentrating risk among those least able to manage it. The structural asymmetries embedded in the ecosystem—fee capture for insiders, mark-to-market risk for retail participants; locked liquidity for promoters, forced liquidations for leveraged positions—ensured that when reversal came, the impact would be distributed inversely to power and information access.

For the millions of retail investors who suffered substantial losses, many of them supporters who had invested based on political conviction rather than financial analysis, the lesson proved difficult. Political affinity does not protect against market forces. Loyalty cannot substitute for diversification. Conviction untempered by risk management leads to loss regardless of how deeply held the underlying beliefs. The $12 billion in losses among $TRUMP holders alone represents not merely financial destruction but a breach of trust between political leaders and followers that may have consequences extending beyond markets.

For economists, regulators, and policymakers, the episode offers critical insights into how emergent asset classes intersect with political authority in the digital age. It illustrates the challenges of unbridled speculative enthusiasm meeting concentrated political power. It exposes structural fragilities that emerge when markets become dominated by sentiment-driven capital flows rather than fundamental value assessment. It demonstrates the need for governance frameworks that can address conflicts of interest, information asymmetries, and systemic risks in politically linked digital assets without stifling legitimate innovation.

The monetary story is ultimately one of both creation and destruction, of fortunes made and lost, of a financial instrument that served simultaneously as wealth generation mechanism for sophisticated insiders and wealth transfer mechanism extracting savings from ordinary participants. Digital currencies proved capable of functioning as both extraordinary wealth generators and instruments of financial stress when aligned with political power. The dual nature of these assets—their capacity for concentrated enrichment and their potential to trigger rapid systemic stress—represents a defining challenge for the next era of financial regulation and monetary policy.

In an era where money, media, and political influence have become inseparably intertwined, where attention is collateral and conviction is capital, and where presidential communications move markets more powerfully than central bank pronouncements, vigilance is essential. Diversification is not merely prudent but necessary. Risk management is required for survival. These are lessons that this event has taught with clarity.

The Trump cryptocurrency saga will be remembered as the moment when the fusion of political power and digital finance reached its extreme—and when the limits of that fusion became apparent. Whether future administrations and market participants learn from this episode or repeat it in new forms remains to be seen. What is certain is that the monetary landscape has been permanently altered, the boundaries between politics and finance irrevocably redrawn, and the understanding of risk in the digital age fundamentally reshaped by the events of 2025.

# Appendix: Supplementary Tables

**Table A1.** Timeline of Key Events

| Date/Period | Event | Significance |
|---|---|---|
| 2024 (Campaign) | Trump adopts pro-crypto stance | Political positioning established |
| Early 2025 | $TRUMP and Melania token launches | Entry into memecoin market |
| Early 2025 | World Liberty Financial debut | DeFi platform expansion |
| Sept 2024 | American Bitcoin Nasdaq listing | Mining operations entry |
| Q4 2024 | Trump Media BTC treasury strategy | Corporate transformation |
| Nov 2024 | $TRUMP reaches $9.49 peak | Political celebration drives value |
| Jan-Jun 2025 | $800M crypto sales income | Peak extraction period |
| Oct 2025 | Bitcoin reaches $126,000 ATH | Market peak |
| Oct 10, 2025 | 100% China tariff announcement | Crash trigger |
| Oct 10-11, 2025 | $19B liquidations, 1.6M positions | Cascade phase |
| Nov 2025 | Bitcoin falls below $82,000 | Trough period |
| Nov 2025 | Family wealth: $7.7B to $6.7B | Wealth impact |

**Table A2.** Glossary of Technical Terms

| Term | Definition |
|---|---|
| Altcoin | Any cryptocurrency other than Bitcoin |
| Circuit Breaker | Trading halt mechanism during extreme moves; absent in crypto |

| Term | Definition |
|---|---|
| DeFi | Decentralized Finance—blockchain-based financial services |
| Leverage | Borrowed capital amplifying gains and losses |
| Liquidation | Forced closure of leveraged position |
| Memecoin | Cryptocurrency valued by cultural/celebrity association |
| Oracle | System providing external data to blockchain contracts |
| On-Chain | Data recorded on public blockchains |
| Treasury Strategy | Corporate policy of holding cryptocurrency assets |